# An Improved Multi-Agent Algorithm for Cooperative and Competitive Environments by Identifying and Encouraging Cooperation among Agents

JunjieQi[1], Siqi MAO[2], and Tianyi TAN[3]

[1]Department of Industrial Engineering, Nanjing University, Nanjing, 210046,China, 201870132@smail.nju.edu.cn
[2]Department of Mathematics, Dalian University of Technology, Panjin, 124000, China, maodalian6@mail.dlut.edu.cn
[3]Department of Mathematics, Dalian University of Technology, Panjin, 124000, China, tantianyi@mail.dlut.edu.cn

December 1, 2024

## Abstract

We propose an improved algorithm by identifying and encouraging the cooperative behavior in the multi-agent environment. First, we analyze the shortcomings of existing algorithms to address multi-agent reinforcement learning problem. Then, based on the existing algorithm MADDPG, we employ a new parameter to increase the reward that an agent can get when we identify an cooperative behavior occurs among all agents. Finally, We compare our improved algorithm and MADDPG in the environment from pettingzoo. The results show that the new algorithm can help agents obtain both higer team reward and individual reward.

# 1 Introduction

In today's technological world, the coordination and cooperation of multiple agents play a crucial role in various real-world applications. For instance, in autonomous vehicle fleets navigating city streets or drone swarms conducting search-and-rescue missions, the synchronization of multiple agents is essential. Take a fleet of delivery drones as an example: each drone must avoid collisions and optimize its path in harmony with others to ensure timely deliveries. These examples vividly illustrate the importance of teamwork in achieving successful outcomes in multi-agent systems.Unlike single-agent reinforcement learning, where the focus is on finding an optimal policy for a single agent, the multi-agent scenario introduces significant complexity[1].

In this study, we address the challenges of developing effective multi-agent reinforcement learning (MARL) algorithms that can facilitate cooperation among agents in both competitive and cooperative environments. Traditional reinforcement learning (RL) algorithms may fail in multi-agent settings due to the environment's non-stationary nature, which violates the Markov assumption inherent in Q-learning[2]. In these environments, the actions of each agent affect the state, making it dynamic and unpredictable, thus complicating the learning process[3]. Policy gradient methods, while effective in single-agent RL, face computational difficulties as the number of agents increases, leading to higher variance and slower convergence[4]. Recent advancements, such as Multi-Agent Deep Deterministic Policy Gradient (MADDPG)[5], address these issues by enabling agents to learn and coordinate effectively, thereby approximating a Markov process and improving learning stability.

To overcome these challenges, we propose an advanced MARL approach that builds on existing methods like Multi-Agent Deep Deterministic Policy Gradient (MADDPG) to predict other agents' policies, thereby approximating a Markov process. Because our environment operates in a continuous space, which is not suitable for value iteration using Deep Q-Network (DQN), we also employ the Proximal Policy Optimization (PPO) method[6].

# 2  Background

We consider the same reinforcement learning scenario, involving multiple agents interacting within a shared environment E over discrete time steps. At each time step teach agent i receives an observation $o_t^i$ from the environment, takes an action $a_t^i$ [7], and subsequently receives a reward $r_t^i$. In such environments, the observations, actions, and rewards are influenced by the collective behavior of all agents, making the problem inherently more complex compared to single-agent scenarios.

The behavior of each agent follows a policy $\mu^i$. The environment E can be represented as a Markov Game, defined by the tuple $\left(\mathcal{N}, S, \{A^i\}_{i=1}^N, \{o^i\}_{i=1}^N, P, \{r^i\}_{i=1}^N\right)_i$ [8], where N is the set of agents, S is the state space, $A_i$ is the action space of agent i, $o_i$ is the observation space of agent i, P is the state transition function, and $r_i$ is the reward function for agent[9].

In cooperative or competitive environments, the actions of each agent impact the state transition probabilities, often leading to non-stationary dynamics from the perspective of any single agent[10]. This non-stationarity violates the Markov assumption traditionally used in rein- forcement learning, where the environment's dynamics are assumed to be fixed. The goal in multi-agent reinforcement learning (MARL) is to find a set of policies $\{\mu^i\}_{i=1}^N$ that optimize the expected cumulative reward for each agent[11].

The value function in MARL is represented by the Q-function[12], which describes the expected cumulative reward for an agent i given a state $s_t$, an action $a_t^i$, and assuming all agents follow theirrespective policies [13]: $Q^{\mu^i}(s_t, a_t^i) = E\left[\sum_{k=t}^{T} \gamma^{k-t} r_k^i \mid s_t, a_t^i, \mu^i\right]$

For deterministic policies, the Q-function can be updated using the Bellman equation[14]:

$$Q^{\mu^i}\left(s_t, a_t^i\right) = E\left[r_t^i + \gamma \max_{a_t^i} Q^{\mu^i}\left(s_{t+1}, a_{t+1}^{i'}\right) \mid s_t, a_t^i\right] \qquad (1)$$

Policy gradient methods are a class of algorithms that directly optimize the policy parameters by calculating the gradient of the expected return with respect to the policy parameters.

Unlike value-based methods, such as Q-learning[15], policy gradient methods directly search for the optimal policy without explicitly representing the value function. This approach can be particularly effective in complex environments where estimate an accurate value is challeng- ing.

Traditional reinforcement algorithms like Q-learning[16] and policy gradient methods have limitations when applied to MARL settings. Q-learning struggles to estimate the Q-value accurately due to the changing dynamics caused by other agents' actions, while policy gradient methods face difficulties in computing gradients in such environments. Recent advancements, such as the Multi-Agent Deep Deterministic Policy Gradient (MADDPG) algorithm[17], address challenges by incorporating centralized training with decentralized execution. This approach allows agents to predict other agents' policies, treating the multi-agent environment as if it were closer to a Markov process, thus stabilizing the learning process.

To further enhance cooperation in multi-agent environments, improvements such as incorporating team rewards have been proposed. These approaches focus on rewarding agents not only for their individual achievements but also for contributing to the success of the group, fostering a higher level of collaboration among agents.

## 3 Methodology

In this section, we aim to derive an improved algorithm that can enable each agent to exhibit better cooperative behavior. As our algorithm is based on the MADDPG algorithm, We will operate under the same constraints in MADDPG: (1) do not assume any particular structure on the communication method between agents (2) do not assume a differentiable model of the environment dynamics. We will improve the MADDPG algorithm by increasing the reward that each agent can receive in some situations to encourage the cooperative behavior.

We used the same notation as in Ryan Lowe's paper[7]. In our game, there is a competitive and cooperative relationship among N agents consisting of two teams. Each agent's policy is parameterized by $\theta = \{\theta_1,..., \theta_N\}$, and the set of all agent policies is parameterized by $\pi = \{\pi_1,..., \pi_N\}$. In the background section, notation used in MADDPG has been introduced clearly. Based on this existing algorithm, we introduced a new parameter $\varphi_i$ to reward agent i if there exists any cooperative behavior. Then, we can updated the value function $Q_i^\mu$ for agent i as follows:

$$L(\theta_i) = E_{x,a,r,x'}\left[(Q_i^\mu(x, a_1,..., a_N) - y)^2\right], y = \varphi_i r_i + \gamma Q_i^{\mu'}(x', a_1',..., a_N')\Big|_{a_j' = \mu_j'(o_j)} \quad (2)$$

We will introduce the values of $\varphi_i$ in detail. Let m be the number of agents in agent i's team. Without loss of generality, we can represent the rewards of all agents in the agent i's team by the set $\{r_1,..., r_m\}$. For the set $\{r_1,..., r_m\}$, we can easily calculate the total number k of positive reward in this set. Then, we can calculate $\varphi_i$ as follows:

$$\phi_i = \begin{cases} \phi, & k > L \\ 1, & \text{othewise} \end{cases} \quad (3)$$

L and ϕ are predetermined hyperparameters. L takes natural number values and represents the minimum number of agents required to cooperate. When L is small, it indicates that we believe there can exist cooperation with only a few agents having positive rewards. Conversely, when L is large, it means that we consider cooperation to occur only if many agents have positive rewards. ϕ is a positive real number and represents the extent to which we encourage cooperative behavior. The greater the value of ϕ, the more additional rewards agent i can obtain. In other words, as ϕ increases, agent i can achieve higher rewards through contribution to cooperative behaviors. This, in turn, can promote more intense cooperative behaviors across our environment. Therefore, formulation (2) means when k exceeds L, agent i will get a higher reward **φ**r$_i$ instead of r$_i$. The process of our improved algorithm is shown in Algorithm 1.

A primary motivation behind our algorithm is that we think the individual rewards of most of the cooperating agents are positive when agents exhibit cooperative behavior. Upon identifying such situations, we strengthen the learning of cooperative behavior by increasing the overall reward resulting from cooperation during training through the new parameter **φ**$_i$.

---

**Algorithm 1:** The improved algorithm based on MADDPG with the new parameter $\phi_i$

**Input:** Number of episodes M, max episode length, discount factor γ, minibatch size S

1. **for** *episode = 1 to M* **do**
2.    Initialize a random process N for action exploration;
3.    Obtain initial state **x**;
4.    **for** t = 1 *to maximum episode length* **do**
5.      **for** *each agent* i **do**
6.        Choose action a$_i$ = **μ**$_{θ_i}$(o$_i$) + N$_t$ based on the current policy and exploration;
7.      Execute actions a = (a$_1$,..., a$_N$) and observe reward r and new state **x**′;
8.      Store the tuple (**x**,a,r,**x**′) in replay buffer D;
9.      Update the state **x** ← **x**′;
10.      **for** *each agent* i = 1 *to* N **do**
11.        Sample a random minibatch of S experiences (**x**$^j$, a$^j$, r$^j$, **x**′$^j$) from D;
12.        Calculate $\phi_i$ according to the formula (3);
13.        Set $y^j = \phi_i r_i^j + \gamma Q_i^{\mu'}(\mathbf{x}'^j, a_1', \ldots, a_N')\Big|_{a_k' = \boldsymbol{\mu}_k'(o_k^j)}$;
14.        Update the critic by minimizing the loss:

$$\mathcal{L}(\theta_i) = \frac{1}{S} \sum_j \left(y^j - Q_i^{\mu}(\mathbf{x}^j, a_1^j, \ldots, a_N^j)\right)^2$$

          Update actor using the sampled policy gradient:

$$\nabla_{\theta_i} J \approx \frac{1}{S} \sum_j \nabla_{\theta_i} \boldsymbol{\mu}_i(o_i^j) \nabla_{a_i} Q_i^{\mu}(\mathbf{x}^j, a_1^j, \ldots, a_i, \ldots, a_N^j)\Big|_{a_i = \boldsymbol{\mu}_i(o_i^j)}$$

15.      Update target network parameters for each agent i:

$$\theta_i' \leftarrow \tau\theta_i + (1 - \tau)\theta_i'$$

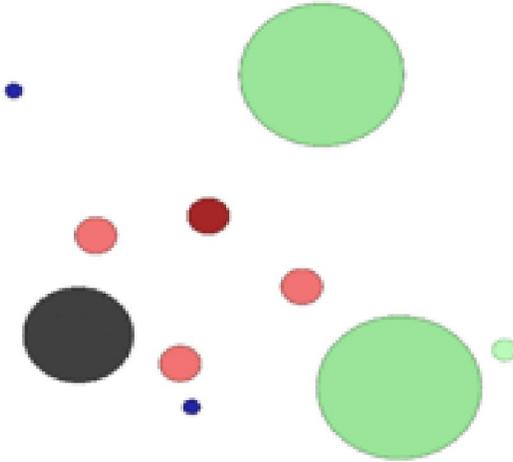

Figure 1: The MPE environment

# 4 Experiments

Our experiment is focused on the comparison between MADDPG and our algorithm. We used the hyperparameters L and φ with values of 1 and 2, respectively. For the other parameters, our algorithm adopts the same default parameter settings as the MADDPG algorithm.

Our environment is Multi-Particle Environments (MPE) from PettingZoo, as shown in Figure 1. We have six agents consisting of four red agents and two green agents, and three obstacles. The action space is discrete, which involves up, down, left, and right. When the red agent get close or catch the green one, they will have reward. When the green one get closer to the water, they will be rewarded too. To evaluate our algorithm's performance, We compare the total reward of the red team in this environment and the reward of each agent separately. We plot the total reward convergence over 25000 episodes in Figure 2 and 3. We observe that our algorithm outperforms MADDPG in terms of the total reward of all agents. For the total reward of all agents in both the red team and the green team, our algorithm has higher reward than the MADDPG. We also compare each team's reward in Fig 1. For the green team, training results of the two algorithms are almost the same. However, the red agents in our algorithm will get a higher reward, which means our algorithm helps red agents find a better policy to catch the green agents.

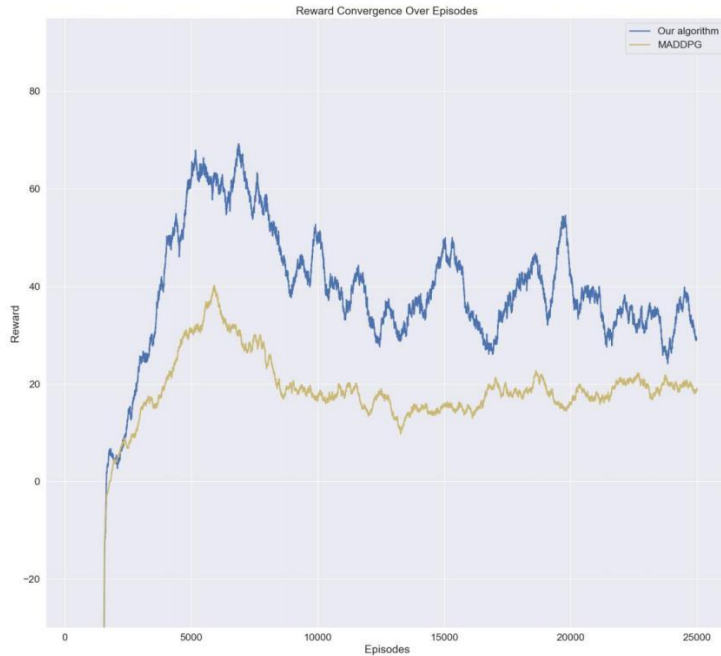

Figure 2: Comparison of MADDPG and our algorithm:plot of two team's total reward

Besides, we plot the reward change of each agent in red team in Figure 4. In the early stages of training, the rewards received by each agent vary. However, as training progresses, the trends in the rewards for each agent become very similar. Our algorithm's agents in the red team outperform the MADDPG agents by achieving higher individual rewards.

# 5  Conclusions

Based on the existing algorithm MADDPG, we introduce a new parameter $\varphi_i$ to encourage the cooperative behavior between agents in the same team. We have tested our new algorithm in a both cooperative and competitive environment, consisting of two teams. The empirical experiments show that the improved algorithm outperforms MADDPG in terms of the final

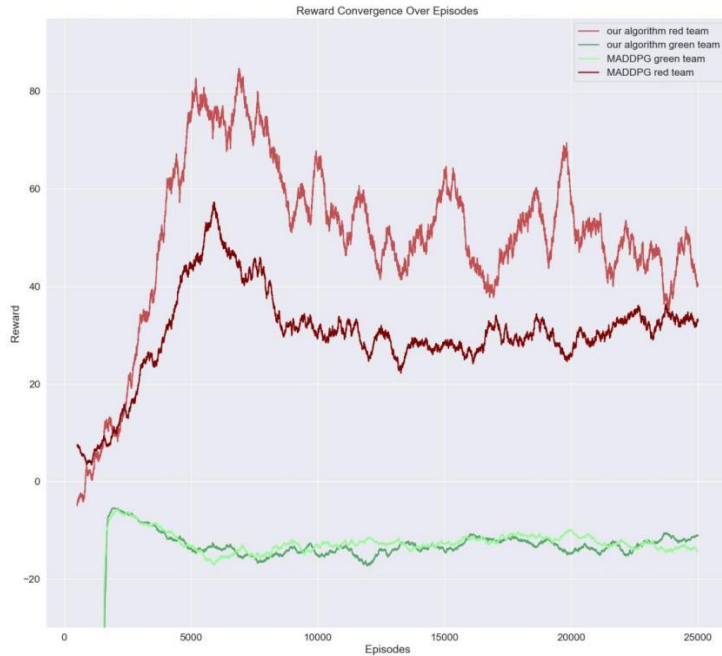

Figure 3: Comparison of MADDPG and our algorithm:plot of each team's reward reward.

The result indicates that the improved algorithm can find a better policy to help agents in same team cooperate better. Under this policy, the agent's team gets a higher team reward and the agent's individual reward increases to a large extent as well.

# References

[1] Tan, M.: Multi-agent reinforcement learning: independent vs. cooperative agents. In: Proceedings of the Tenth International Conference on Machine Learning, pp. 330–337. Morgan Kaufmann (1993)

[2] Canese, L., et al.: Multi-agent reinforcement learning: a review of challenges and applications. Appl. Sci. 11(11) (2021).

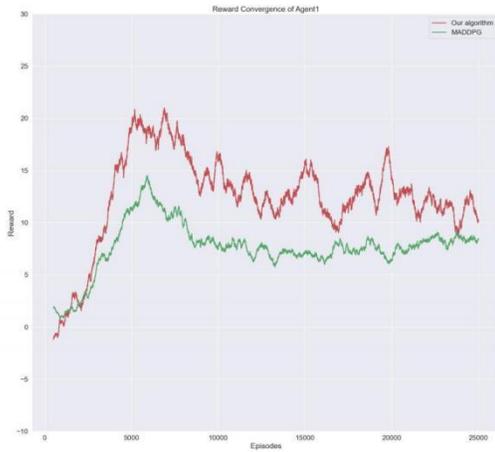
(a) Agent1

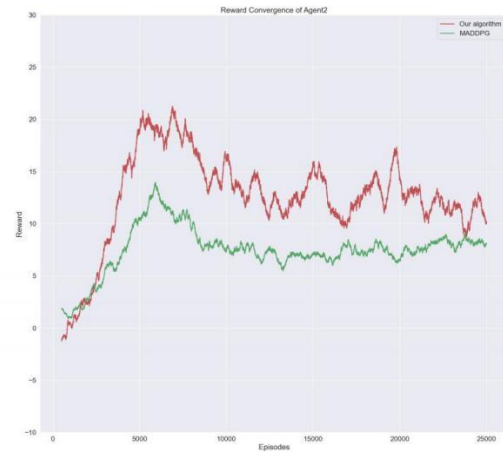
(b) Agent2

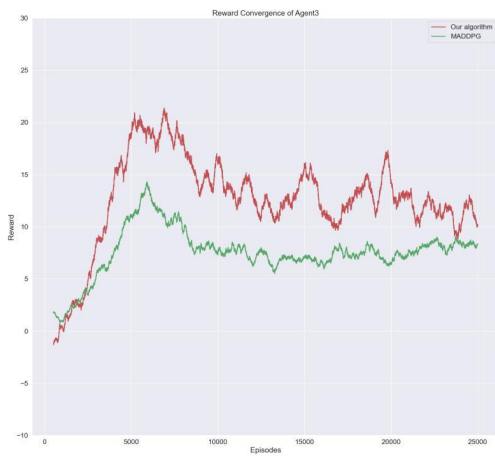
(c) Agent3

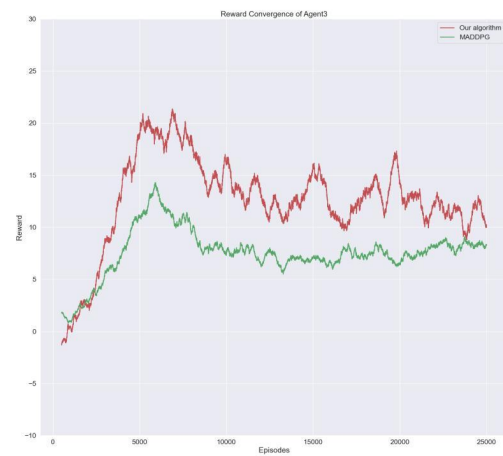
(d) Agent4

Figure 4: Comparison of MADDPG and our algorithm on each agent's reward.

[3] Papoudakis, G., Christianos, F., Rahman, A., Albrecht, S.V.: Dealing with non-stationarity in multi-agent deep reinforcement learning. CoRR abs/1906.04737 (2019).

[4] Daskalakis, C., Foster, D.J., Golowich, N.: Independent policy gradient methods for competitive reinforcement learning. Adv. Neural Inf. Process. Syst. 33, 5527–5540 (2020)

[5] Samende, C., Cao, J., Fan, Z. (2022). Multi-agent deep deterministic policy gradient algorithm for peer-to-peer energy trading considering distribution network constraints.

Applied Energy, 317, 119123. https://doi.org/10.1016/j.apenergy.2022.119123

[6] Schulman, J., Wolski, F., Dhariwal, P., Radford, A., Klimov, O. (2017). Proximal Policy Optimization Algorithms. arXiv preprint arXiv:1707.06347. Retrieved from

[7] Lowe R, Wu Y I, Tamar A, et al. Multi-agent actor-critic for mixed cooperative-competitive environments[J]. Advances in neural information processing systems, 2017, 30.

[8] Oliehoek F A, Amato C. A Concise Introduction to Decentralized POMDPs[M]. Springer, 2016.

[9] Gupta J K, Egorov M, Kochenderfer M. Cooperative multi-agent control using deep reinforcement learning[C]//International Conference on Autonomous Agents and Multiagent Systems. 2017: 66-83.

[10] Foerster J, Assael Y M, de Freitas N, et al. Learning to communicate with deep multi-agent reinforcement learning[C]//Advances in Neural Information Processing Systems. 2016, 29.

[11] Yang M, Liu G, Zhou Z. Partially Observable Mean Field Multi-Agent Reinforcement Learning Based on Graph-Attention[J]. arXiv preprint arXiv:2304.12653, 2023.

[12] Duan X, Zhang Y, Wang H, et al. Multi-agent reinforcement learning: A review of challenges and applications[J]. IEEE Access, 2019, 7: 101996-102016.

[13] Zhang C, Lesser V. Coordinating multi-agent reinforcement learning with limited communication[C]//Proceedings of the 2013 international conference on Autonomous agents and multi-agent systems. 2013: 1101-1108.

[14] Rashid T, Samvelyan M, de Witt C S, et al. QMIX: Monotonic value function factorisation for deep multi-agent reinforcement learning[C]//International Conference on Machine Learning. PMLR, 2018: 4295-4304.

[15] Mahajan A, Samvelyan M, Mao L, et al. Tesseract: Tensorised Actors for Multi-Agent Reinforcement Learning[C]//International Conference on Machine Learning. PMLR, 2021: 7313-7324.

[16] Yu C, Yang X, Gao J, et al. Asynchronous Multi-Agent Reinforcement Learning for Efficient Real-Time Multi-Robot Cooperative Exploration[J]. arXiv preprint arXiv:2301.03398, 2023.

[17] Harmer J, Gisslén L, del Val J, et al. Imitation Learning with Concurrent Actions in 3D Games[J]. arXiv preprint arXiv:1803.05402, 2018.

## Acknowledgement

Junjie Qi, Siqi MAO, and Tianyi TAN contributed equally to this work and should be considered co-first authors.